\documentclass[aps,prc,preprint,amsmath,showpacs,superscriptaddress,nofootinbib]{revtex4-1}
\usepackage{graphicx,epstopdf,epsfig,hyperref,color}

\begin{document}
	
	\title{Experimental constraints on shallow heating in accreting neutron-star crusts}
	
	\author{N. Chamel}
	\affiliation{Institut d'Astronomie et d'Astrophysique, CP-226, Boulevard du Triomphe, Universit\'e Libre de Bruxelles, 1050 Brussels, Belgium}
	
	\author{A.~F. Fantina}
	\affiliation{Grand Acc\'el\'erateur National d'Ions Lourds (GANIL), CEA/DRF -
		CNRS/IN2P3, Boulevard Henri Becquerel, 14076 Caen, France}
	\affiliation{Institut d'Astronomie et d'Astrophysique, CP-226, Boulevard du Triomphe, Universit\'e Libre de Bruxelles, 1050 Brussels, Belgium}
	
	\author{J.~L. Zdunik}
	\affiliation{N. Copernicus Astronomical Center, Polish Academy of Sciences, Bartycka 18, PL-00-716, Warszawa, Poland}
	
	\author{P. Haensel}
	\affiliation{N. Copernicus Astronomical Center, Polish Academy of Sciences, Bartycka 18, PL-00-716, Warszawa, Poland}

	\begin{abstract}
The observed thermal relaxation of transiently accreting neutron stars during quiescence periods in low-mass X-ray binaries suggests the existence of unknown heat sources in the shallow layers of neutron-star crust. Making use of existing experimental nuclear data, we estimate the maximum possible amount of heat that can be deposited in the outer crust of an accreting neutron star due to electron captures and pycnonuclear fusion reactions triggered by the burial of X-ray burst ashes. 
	\end{abstract}

	\maketitle

\section{Introduction}
\label{sec:intro}

The accretion of matter onto a neutron star in a low-mass X-ray binary induces various
nuclear processes leading to remarkable astrophysical phenomena~\cite{parikh2013,galloway,degenaar2018}. In these systems, matter
is transferred from a low-mass stellar companion to a neutron star via an accretion disk. The
hydrogen-rich material that accumulates at the surface of the neutron star burns steadily producing a thick helium layer. Once the critical conditions for helium ignition are reached (typically at density of order $10^7$ g cm$^{-3}$), the overlying envelope is converted
into heavier nuclides within seconds. These thermonuclear explosions are observed as X-ray
bursts, with luminosity up to about $10^{38}~{\rm erg~s^{-1}}$ (corresponding approximately to the Eddington limit), and with a typical decay lasting a few tens of seconds. The X-ray bursts  are quasiperiodic, with typical recurrence time of about hours to days. Less frequent but
more energetic are superbursts lasting for a few hours with recurrence timescales of several
years. With a total energy of order $ 10^{42}$~erg, superbursts are likely to be triggered by 
the unstable carbon burning in deep layers of the outer crust. 

In most X-ray binaries, accretion is
not persistent but occurs sporadically. In particular, soft X-ray transients (SXTs) exhibit active
periods of weeks to months separated by quiescence periods of years to decades during which there is very little or no accretion. So-called
‘quasipersistent’ SXTs remain active for years to decades. As matter accumulates on the neutron-star
surface, ashes of X-ray bursts are buried into deeper layers and further processed (see, e.g., Ref.~\cite{meisel2018} for a recent review). These reactions release about 1.5-2 MeV per accreted nucleon in the inner crust (see, e.g. Refs.~\cite{lau2018,fantina2018,shchechilin2019} and references therein). During quiescence periods, the heat is transported to the surface thus cooling down the star. In quasipersistent SXTs, the
accretion can last long enough for the crust to be driven out of thermal equilibrium with the core. X-ray observations thus provide unique probes of the neutron-star crust properties~\cite{lrr}. 

The evolution of the effective surface temperature of a
few quasipersistent SXTs during quiescence has been inferred from observations of their thermal emission~\cite{wijnands2017}. The thermal relaxation of the crust is still
not fully understood. In particular, the interpretation of the cooling data requires the existence of
some unknown heat sources in the shallow layers of the crust~\cite{brown2009,degenaar2011,degenaar2013,degenaar2014,homan2014,degenaar2015,turlione2015,deibel2015,merritt2016,ootes2016,waterhouse2016,degenaar2017,parikh2017,parikh2018,ootes2018,ootes2019,degenaar2019,parikh2019}, as summarized in Table~\ref{tab:heating}. In most cases, the extra heat, typically released in the outer crust at densities below about $10^{10}$~g~cm$^{-3}$, amounts to about 1-2 MeV per accreted nucleon with the notable exceptions of MAXI J0556$-$332 and Aql X-1. Part of this shallow heating could be explained by uncertainties in the accretion rate~\cite{degenaar2014,turlione2015,ootes2016} and the 
envelope composition~\cite{ootes2018}. The latent heat accompanying the crystallization of the crust represents only a few keV at most (see e.g. Section 2.3.4 in Ref.~\cite{haensel2007}). Compositionally driven convection in the ocean could account for about 0.2 MeV per nucleon~\cite{medin2011,medin2014,medin2015}. Differential rotation between the liquid ocean and the solid crust could be another source of heating~\cite{inogamov2010}. 

In this paper, we study the maximum amount of heat that can possibly be deposited in the shallow layers of the outer crust of an accreting neutron star, from electron captures by nuclei and pycnonuclear fusion reactions making use of the available experimental nuclear data.

\begin{table}
	\centering
	\caption{Heat per one accreted nucleon (in MeV) deposited in the shallow layers of accreting neutron-star crusts, as required by cooling simulations to fit the observational data. The indicated amount of heat may correspond to different accretion periods.}
	\label{tab:heating}
	\vspace{.5cm}
	\begin{tabular}{|cc|}
		\hline 
		Swift J174805.3$-$244637 & 1.4 \cite{degenaar2015} \\
		XTE J1701$-$462 & 0.17 \cite{turlione2015} \\
		EXO 0748$-$676 & 0.35-1.8 \cite{degenaar2014} \\
		MXB 1659$-$29 & 0.8\cite{brown2009}, 1-1.2\cite{parikh2019}, 1.6\cite{turlione2015} \\ 
		KS 1731$-$260 & 1.38$\pm$0.18 \cite{merritt2016,ootes2016} \\
		Aql X-1 & 0.9-3.7 \cite{waterhouse2016,ootes2018}, 2.3-9.2 \cite{degenaar2019} \\ 
		IGR J17480$-$2446 & 1\cite{degenaar2011,degenaar2013,ootes2019}-3.8\cite{turlione2015}  \\ 
		1RXS J180408.9$-$342058 & 0.9 \cite{parikh2018} \\ 
		MAXI J0556$-$332 & 6-17\cite{homan2014,deibel2015,parikh2017} \\ 
		HETE J1900.1$-$2455 & 0-3\cite{degenaar2017} \\
		\hline 
		\end{tabular} 
\end{table}

\section{Microscopic model of accreted neutron-star crusts}
\label{sec:micro}

\subsection{Compression of X-ray burst ashes}

During accretion, X-ray burst ashes, produced at densities $\rho \lesssim  10^{7}$~g~cm$^{-3}$, sink deeper and deeper under the weight of continuously accumulated material. 
Let us consider the compression of a layer composed of nuclei $(A,Z)$ with proton number $Z$ and mass number 
$A$. Although the crust of accreting neutron stars is expected to contain a mixture of various nuclear species (see, e.g. \cite{lau2018,shchechilin2019}), the amount of heat deposited by nuclear processes, which is the main focus of this work, was shown to be well reproduced by the single-nucleus approximation~\cite{haensel2008,fantina2018,shchechilin2019}. At the densities of interest ($\rho> 10^8$ g~cm$^{-3}$), atoms are fully ionized by the tremendous pressure~\cite{haensel2007}. For temperatures of order $10^8$~K, the shallowest regions of accreted neutron-star crusts 
remain liquid. Crystallization to a body-centered cubic lattice occurs at temperature $T_m$, given by~\cite{haensel2007}
\begin{equation}\label{eq:Tm}
T_m=\frac{e^2}{a_e k_\text{B} \Gamma_m}Z^{5/3}\approx 1.3\times 10^5 Z^2 \left(\frac{\rho_6}{A}\right)^{1/3}~{\rm K}\, ,
\end{equation}
$e$ being the elementary electric charge, $a_e=(3/(4\pi n_e))^{1/3}$ the electron-sphere radius, $n_e$ 
the electron number density, $k_\text{B}$ Boltzmann's constant, $\Gamma_m\approx 175$ the Coulomb 
coupling parameter at melting, and $\rho_6=\rho/10^6$~g~cm$^{-3}$. At the lowest densities considered here, the temperature $T$ is typically of the same order as $T_m$ so that the Coulomb liquid is very close to the crystallization transition, and is strongly coupled to the extent that the 
temperature $T$ is much smaller than $T_\ell$ defined by~\cite{haensel2007} 
\begin{equation}
T_\ell = \frac{Z^2 e^2}{a_i k_\text{B}} \approx 2.3\times 10^7 Z^2 \left(\frac{\rho_6}{A}\right)^{1/3}~{\rm K}\, ,
\end{equation}
where $a_i=(3/(4\pi n_N))^{1/3}$ is the ion-sphere radius and $n_N$ is the number density of ions (bare atomic nuclei). 
Because the temperature $T$ is much lower than the electron Fermi temperature defined by 
\begin{equation}\label{eq:TFe}
T_{\text{F}e}=\frac{\mu_e-m_e c^2}{k_\text{B}}\approx 6.0\times 10^9 \left(\frac{Z}{A}\rho_6 \right)^{1/3}~{\rm K}\, ,
\end{equation}
where $\mu_e$ is the electron Fermi energy, $m_e$ the electron mass, and $c$ is the speed of light, electrons are highly 
degenerate. In what follows, the thermal contributions to the thermodynamic potentials will be neglected, and calculated at $T=0$~K. 

Because matter is ``relatively cold'', thermonuclear processes are strongly suppressed. 
However, nuclei may capture electrons with 
the emission of neutrinos in two steps: 
\begin{equation}
(A,Z)+e^-\longrightarrow  (A,Z-1)+\nu_e \, ,
\label{eq.sect.accretion.processes.ecap1}
\end{equation}
\begin{equation}
(A,Z-1)+e^-\longrightarrow  (A,Z-2)+\nu_e + q \, .
\label{eq.sect.accretion.processes.ecap2}
\end{equation}
The first capture proceeds in \emph{quasi-equilibrium}, with negligible energy release. The daughter nucleus is generally highly unstable, 
and captures a second electron \emph{off equilibrium} with an energy release $q$. 
In hydrostatic equilibrium, the pressure throughout the star must vary continuously so that this process 
occurs at a fixed pressure $P_\beta$. Since the temperature is also fixed, and the nucleon number is conserved, 
the suitable thermodynamic potential for determining the equilibrium composition of the crust is the Gibbs free energy per nucleon $g$~\cite{tondeur1971,bps1971}. 
In particular, a nucleus will be unstable against electron captures if the Gibbs free energy per nucleon is  
higher than that of its daughter nucleus. The reaction (\ref{eq.sect.accretion.processes.ecap1}) will generally involve a transition 
from the parent nucleus ground state to the lowest excited state of the daughter nucleus, as determined by Fermi and Gamow-Teller 
selection rules. However, if the parent nucleus sinks slowly enough as compared to the timescale of electron captures, a transition 
to the daughter nucleus ground state may occur at a lower pressure. We shall consider these two scenarios. 
At high enough densities, light nuclei such as carbon and oxygen may fuse \cite{horowitz2008,lau2018}. We shall also study such pycnonuclear reactions. 
As nuclei sink deeper into the crust, they become so neutron rich that they are unstable against neutron emission~\cite{chamel2015,fantina2016a}. After capturing an electron, the nucleus $(A,Z)$ will thus transform into a nucleus $(A-\Delta N,Z-1)$ with the emission of $\Delta N$ neutrons $n$ and an electron neutrino $\nu_e$~:
\begin{equation}\label{eq:e-capture+n-emission}
(A,Z)+ e^- \rightarrow (A-\Delta N,Z-1)+\Delta N n+ \nu_e\, .
\end{equation}
The onset of neutron emission delimits the boundary between the outer and inner crusts. Neutrons may also be transferred between nuclei~\cite{chugunov2019}. However, the rates of these reactions remain very uncertain. We shall therefore not consider these reactions.

\subsection{Gibbs free energy per nucleon}

The Gibbs free energy per nucleon at pressure $P$ is defined by
\begin{equation}
\label{eq:gibbs-def}
g=\frac{\mathcal{E}+P}{\bar n}\, ,
\end{equation}
where $\bar n$ is the average nucleon number density, and $\mathcal{E}$ is the average energy density. 
For typical temperatures of order $10^7$~K, we can safely assume that before capturing electrons nuclei are in their ground state. 
Their energy density $\mathcal{E}_N$ is thus given by 
\begin{equation}
\label{eq:EN}
\mathcal{E}_N=n_N M^\prime(A,Z)c^2\, ,
\end{equation}
where $M^\prime(A,Z)$ denotes the nuclear mass (including the rest mass of $Z$ protons, $A-Z$ neutrons and 
$Z$ electrons \footnote{The reason for including the electron rest mass in $M(A,Z)$ is that \emph{atomic} masses are generally 
tabulated rather than \emph{nuclear} masses.}). After the first electron capture, the daughter nucleus may be in an excited state
with an energy $E_{\rm ex}(A,Z)$. To account for such a possibility, we shall therefore introduce the Gibbs free energy per nucleon $g^*$ 
of excited nuclei with the nuclear energy density 
\begin{equation}
\label{eq:ENstar}
\mathcal{E}_N^*=n_N \biggl[M^\prime(A,Z)c^2+E_{\rm ex}(A,Z)\biggr]\, .
\end{equation}
To a very good approximation, electrons can be treated as an ideal relativistic Fermi gas. The expression of the electron energy density $\mathcal{E}_e$ 
and pressure $P_e$ can be found in Chap. 2 of Ref.~\cite{haensel2007}. The main correction arises from 
electrostatic interactions. For a strongly coupled one-component plasma of pointlike ions in a uniform charge compensating background of electrons, 
the electrostatic energy density is of the form (see e.g. Chapter 2 of Ref.~\cite{haensel2007})
\begin{equation}\label{eq:EL}
\mathcal{E}_L=C e^2 n_e^{4/3} Z^{2/3}\, ,
\end{equation}
where $C$ is a dimensionless constant. 
We have omitted here the negligibly small contribution due to (thermal and/or quantum) motion of ions about their equilibrium position~\cite{baiko09}. 
The lattice contribution to the pressure is thus given by  
\begin{equation}\label{eq:PL}
 P_L=n_e^2 \frac{d(\mathcal{E}_L/n_e)}{dn_e}=\frac{\mathcal{E}_L}{3}\, . 
\end{equation}
In the liquid phase, this constant is about $C\approx -1.4621$ (this value was calculated using the constant $A_1$ in Ref.~\cite{haensel2007} page 75 as $C=A_1 (4\pi/3)^{1/3}$)
while its value in the solid 
phase (considering a body-centered cubic lattice) is about $C\approx -1.4442$~\cite{haensel2007}. In the following, we shall adopt the Wigner-Seitz 
estimate~\cite{salpeter54}
\begin{equation}\label{eq:WS-approx}
C=-\frac{9}{10}\left(\frac{4\pi}{3}\right)^{1/3}\approx -1.4508\, .
\end{equation}
This model is further refined by taking into account electron exchange and polarization effects as in Ref.~\cite{cf16}. The first correction is included by rescaling 
the energy density $\mathcal{E}_e$ and the electron pressure $P_e$ by a factor $1+\alpha/(2\pi)$, where $\alpha=e^2/(\hbar c)$ is the 
fine structure constant. Electron polarization effects are included by introducing the effective charge number in the electrostatic contributions
\begin{equation}\label{eq:zeff}
Z_{\rm eff}\equiv Z \sigma(Z)^{3/2}\, ,
\end{equation}
where 
\begin{equation}
\sigma(Z)=1+\alpha \frac{12^{4/3}}{35 \pi^{1/3}} b_1(Z) Z^{2/3}\, ,
\end{equation}
and the function $b_1(Z)$ is given by~\cite{pot00}
\begin{equation}
b_1(Z) = 1 - 1.1866 \,Z^{-0.267} + 0.27\,Z^{-1}\, .
\end{equation}
Although this expression of the electron polarization was obtained for a Coulomb crystal, it still remains essentially the same for a strongly coupled Coulomb 
liquid~\cite{haensel2007}. We shall therefore adopt the same expression in both phases.  Note that the effects of electron charge polarization is to effectively increase the proton number $Z_{\rm eff}>Z$. 
For a discussion of higher-order corrections, 
see e.g. Ref.~\cite{pearson2011}. 
Collecting all terms, and imposing electric charge neutrality $n_e=Z n_N$, the Gibbs free energy~(\ref{eq:gibbs-def}) can thus be written as~\cite{cf16}
\begin{equation}
\label{eq:gibbs}
g=\frac{M^\prime(A,Z)c^2}{A}+\frac{Z}{A}\biggl[\mu_e\left(1+\frac{\alpha}{2\pi}\right) -m_e c^2+\frac{4}{3} C \alpha \hbar c n_e^{1/3} Z_{\rm eff}^{2/3}\biggr]\, .
\end{equation}
where $\mu_e=(\mathcal{E}_e+P_e)/n_e$ is the electron Fermi energy. For nuclei in excited states, the Gibbs free energy per nucleon $g^*$ takes a 
similar form except that $M^\prime(A,Z)c^2$ must be replaced by $M^\prime(A,Z)c^2+E_{\rm ex}(A,Z)$. 
The pressure reads~\cite{cf16}
\begin{equation}
\label{eq:pressure}
 P=P_e\left(1+\frac{\alpha}{2\pi}\right) + \frac{C}{3} e^2 n_e^{4/3} Z_{\rm eff}^{2/3}\, .
\end{equation}
We have neglected the thermal pressure of ions, which is of order $\alpha Z^{2/3} (T/T_\ell)P_e$~\cite{haensel2007}. 
In the regime of ultrarelativistic electrons $\mu_e\gg m_e c^2$, the pressure is approximately given by~\cite{cf16}
\begin{equation}\label{eq:P-ultra-rel}
P\approx \frac{\mu_e^4}{12 \pi^2 (\hbar c)^3}\Biggl[1+ \frac{\alpha}{2\pi}\left(1+\frac{8}{3} C\left(\frac{\pi}{3}\right)^{1/3} Z_{\rm eff}^{2/3}\right) \Biggr] \, .
\end{equation}
Solving for $\mu_e \approx \hbar c (3\pi^2 n_e)^{1/3}$ and substituting in Eq.~(\ref{eq:gibbs}), the Gibbs free energy per nucleon can thus be 
explicitly expressed as a function of $A$, $Z$, and $P$~: 
\begin{equation}
\label{eq:gibbs-ultra}
g\approx \frac{M^\prime(A,Z)c^2}{A}+\frac{Z}{A}\Biggl[1+\frac{\alpha}{2\pi}\left(1+\frac{8}{3} C\left(\frac{\pi}{3}\right)^{1/3} Z_{\rm eff}^{2/3}\right)\Biggr]^{3/4}(12 \pi^2 \hbar^3 c^3 P)^{1/4}\, .
\end{equation}
Let us stress however that this expression is only valid in the ultrarelativistic regime for electrons. 

\subsection{Onset of nuclear processes}

The onset of electron captures by nuclei is determined by the condition $g(A,Z,P_\beta)=g(A,Z-1,P_\beta)$, assuming that the daughter 
nucleus is in its ground state. This condition can be 
expressed to first order in $\alpha$ as~\cite{cf16}
\begin{equation}\label{eq:e-capture-gibbs-approx}
\mu_e\left(1+\frac{\alpha}{2\pi}\right) + C \alpha \hbar c n_e^{1/3}F(Z) =  \mu_e^{\beta} \, ,
\end{equation}
\begin{equation}\label{eq:def-F}
 F(Z)\equiv Z^{5/3}\sigma(Z)-(Z-1)^{5/3}\sigma(Z-1) + \frac{1}{3} Z^{2/3}\sigma(Z)\, ,
\end{equation}
\begin{equation}\label{eq:muebeta}
\mu_e^{\beta}(A,Z)\equiv -Q_{\rm EC}(A,Z) + m_e c^2   \, ,
\end{equation}
where we have introduced the $Q$-value (in vacuum) associated with electron capture by nuclei ($A,Z$) 
\begin{equation}
Q_{\rm EC}(A,Z) = M^\prime(A,Z)c^2-M^\prime(A,Z-1)c^2\, .
\end{equation}
These $Q$-values can be obtained from the tabulated $Q$-values of $\beta$ decay by the following relation
\begin{equation}
 Q_{\rm EC}(A,Z) = -Q_\beta(A,Z-1)\, .
\end{equation}
Note that if $Q_{\rm EC}(A,Z)>0$, the nucleus $(A,Z)$ is unstable against electron captures at any density. Following the analyses of Refs.~\cite{chamel2016,chamel2020} and 
recalling that the electron Fermi energy  is given by 
\begin{equation}
 \mu_e = m_e c^2 \sqrt{1+x_r^2}\, , 
\end{equation}
where  $x_r=\lambda_e k_e$, $\lambda_e=\hbar /(m_e c)$ is the electron Compton wavelength, 
and $k_e=(3\pi^2 n_e)^{1/3}$ is the electron Fermi wave number, the threshold condition~(\ref{eq:e-capture-gibbs-approx}) can be transformed into 
the quadratic polynomial equation
\begin{equation}\label{eq:e-capture-gibbs-approx2}
x_r^2\biggl[\left(1+\frac{\alpha}{2\pi}\right)^2-\tilde F(Z)^2\biggr] + 2 \gamma_e^{\beta} \tilde F(Z) x_r =  (\gamma_e^{\beta})^2-\left(1+\frac{\alpha}{2\pi}\right)^2\, , 
\end{equation}
with 
\begin{equation}\label{eq:gammae-thres}
\gamma_e^{\beta}\equiv \frac{\mu_e^{\beta}}{m_e c^2}\, , 
\end{equation}
\begin{equation}\label{eq:Ftilde}
\tilde F(Z)\equiv \frac{C}{(3\pi^2)^{1/3}}\alpha F(Z)\, .
\end{equation}
Solving Eq.~(\ref{eq:e-capture-gibbs-approx2}) for $x_r$ yields   
\begin{eqnarray}\label{eq:exact-xr}
 x_r=\gamma_e^{\beta} \Biggl\{\left(1+\frac{\alpha}{2\pi}\right)\sqrt{1-\biggl[\left(1+\frac{\alpha}{2\pi}\right)^2-\tilde F(Z)^2\biggr]/(\gamma_e^{\beta})^{2}}
 -\tilde F(Z)\Biggr\} \nonumber \\ 
 \times \Biggl[\left(1+\frac{\alpha}{2\pi}\right)^2-\tilde F (Z)^2\Biggr]^{-1}\, .
\end{eqnarray}
Using Eq.~(\ref{eq:pressure}) and the expression for the ideal electron Fermi gas pressure (see Chapter 2 in Ref.~\cite{haensel2007}), the pressure at 
the onset of electron captures is given by 
\begin{eqnarray}
\label{eq:exact-Pbeta}
P_{\beta}(A,Z)&=&\frac{m_e c^2}{8 \pi^2 \lambda_e^3} \biggl[x_r\left(\frac{2}{3}x_r^2-1\right)\sqrt{1+x_r^2}+\ln(x_r+\sqrt{1+x_r^2})\biggr]\left(1+\frac{\alpha}{2\pi}\right)\nonumber\\
&&+\frac{C \alpha}{3 (3\pi^2)^{4/3}} x_r^4 \frac{m_e c^2}{\lambda_e^3}Z_{\rm eff}^{2/3} \, .
\end{eqnarray}
The corresponding average nucleon number density is given by 
\begin{eqnarray}\label{eq:exact-rhobeta}
\bar n_\beta(A,Z)  = \frac{A}{Z} \frac{x_r^3 }{3 \pi^2 \lambda_e^3}\, .
\end{eqnarray}
The transition is accompanied by a discontinuous change of density given by (see Eqs.~(9) and (10) of Ref.~\cite{chamel2020})
\begin{equation}
 \frac{\Delta \bar n}{\bar n_\beta} = \frac{Z}{Z-1}\biggl[1+\frac{C\alpha}{(3\pi^2)^{1/3}}\left(Z_{\rm eff}^{2/3}-(Z-1)_{\rm eff}^{2/3}\right)\frac{\sqrt{1+x_r^2}}{x_r}\left(1+\frac{\alpha}{2\pi}\right)^{-1}\biggr]-1\, .
\end{equation}
We have made use of the Gibbs-Duhem relation $dP_e=n_e d\mu_e$. 
In the limit of ultrarelativistic electrons, the average threshold baryon density and pressure reduce to the expressions given by~\cite{cf16}
\begin{eqnarray}\label{eq:rhobeta}
 \bar n_\beta(A,Z) \approx \frac{A}{Z} \frac{\mu_e^\beta(A,Z)^3}{3\pi^2 (\hbar c)^3}
 \biggl[1+\frac{\alpha}{2\pi}+\frac{C \alpha}{(3\pi^2)^{1/3}}F(Z)\biggr]^{-3}\, ,
\end{eqnarray}
\begin{eqnarray}\label{eq:Pbeta}
P_\beta(A,Z) \approx \frac{\mu_e^\beta(A,Z)^4}{12 \pi^2 (\hbar c)^3}\biggl[1+\frac{\alpha}{2\pi}+\frac{4C \alpha Z_{\rm eff}^{2/3}}{(81\pi^2)^{1/3}} \biggr] \biggl[1+\frac{\alpha}{2\pi}+\frac{C \alpha}{(3\pi^2)^{1/3}}F(Z)\biggr]^{-4}\, .
\end{eqnarray}
The threshold condition for transitions to nuclei in excited states is $g(A,Z,P^*_\beta)=g^*(A,Z-1,P^*_\beta)$. The corresponding pressure $P^*_\beta$ and 
density $\bar n_\beta^*(A,Z)$ can be obtained from the previous formulas by merely substituting $M^\prime(A,Z-1)c^2$ with $M^\prime(A,Z-1)c^2+E_{\rm ex}(A,Z-1)$. 
Because $E_{\rm ex}(A,Z-1)\geq 0$, such transitions occur at higher pressure $P^*_\beta\geq P_\beta$ and density $\bar n_\beta^*(A,Z)\geq \bar n_\beta(A,Z)$. 

As discussed in Refs.~\cite{chamel2015,fantina2016a}, the first electron capture by the nucleus $(A,Z)$ may be accompanied by the emission of $\Delta N>0$ neutrons. 
The corresponding pressure $P_\textrm{drip}$ and average baryon density $\bar n_\textrm{drip}$ are given by similar expressions as Eqs.~(\ref{eq:exact-Pbeta}) and (\ref{eq:exact-rhobeta}) respectively, except that the threshold electron Fermi energy $\mu_e^\beta(A,Z)$ is now replaced by 
\begin{equation}\label{eq:muedrip-acc}
\mu_e^\textrm{drip}(A,Z)= M^\prime(A-\Delta N,Z-1)c^2-M^\prime(A,Z)c^2+\Delta N m_n c^2 + m_e c^2 \, ,
\end{equation}
assuming that the daughter nucleus is in the ground state. Transitions to excited states can be taken into account by adding the suitable excitation energy  $E_{\rm ex}(A-\Delta N,Z-1)$ to $M^\prime(A-\Delta N,Z-1)c^2$. Neutron emission will thus occur whenever $\mu_e^\textrm{drip}(A,Z) < \mu_e^\beta(A,Z)$.

\section{Shallow heating}

\subsection{Heat sources}

The heat released by electron captures can be determined analytically as follows. The reaction (\ref{eq.sect.accretion.processes.ecap1}) 
will be generally almost immediately followed by a second electron capture (\ref{eq.sect.accretion.processes.ecap2}) on the daughter nucleus 
provided $g(A,Z-2,P_\beta)<g(A,Z,P_\beta)$ or $g(A,Z-2,P^*_\beta)<g(A,Z,P^*_\beta)$ depending on whether the daughter nucleus after the first capture is in its 
ground state or in an excited state respectively. The maximum possible amount of heat deposited in matter per one accreted \emph{nucleus} is given by 
\begin{equation}
\label{eq:heat-released}
\mathcal{Q}(A,Z)=  A\biggl[g(A,Z,P_\beta)-g(A,Z-2,P_\beta)\biggr] 
\end{equation} 
in the first case, and 
\begin{equation}
\label{eq:heat-released-ex}
\mathcal{Q}^*(A,Z)=  A\biggl[g(A,Z,P^*_\beta)-g(A,Z-2,P^*_\beta)\biggr] \, ,
\end{equation} 
in the second case. The corresponding amounts of heat per one accreted \emph{nucleon} are given by $q(A,Z)\equiv \mathcal{Q}(A,Z)/A$ and $q^*(A,Z)\equiv\mathcal{Q}^*(A,Z)/A$ respectively. These estimates represent upper limits since part of the energy is radiated away by neutrinos. Neutrino losses associated with  second electron 
capture, Eq.(\ref{eq.sect.accretion.processes.ecap2}),   and  additional heating resulting from the gamma deexcitation  of a daughter nucleus,  were first studied in Ref.~\cite{BisnoSeidov1970} in the context of white dwarfs. With obvious modifications, the same method can be applied  to  the  electron captures in the outer crust of neutron stars~\cite{BisnoBook2001}. 

Let us first consider ground-state to ground-state
transitions. The two successive electron captures are accompanied by a small discontinuous change $\delta n_e$ of the electron density $n_e$. Requiring the pressure to remain fixed $P_\beta(A,Z)=P(A,Z-2)$ leads to 
\begin{equation}\label{eq:electron-discontinuity}
    \delta n_e \approx \frac{C\alpha\hbar c}{3}\biggl[Z_{\rm eff}^{2/3}-(Z-2)_{\rm eff}^{2/3}\biggr]\frac{dn_e}{d\mu_e}n_e^{1/3} \left(1+\frac{\alpha}{2\pi}\right)^{-1}\, ,
\end{equation}
where we have made use of Eq.~(\ref{eq:pressure}) and the Gibbs-Duhem relation $dP_e=n_e d\mu_e$. Expanding the electron Fermi energy $\mu_e(n_e+\delta n_e)$ to first order in $\delta n_e/n_e$, substituting in the expression of $g(A,Z-2,P_\beta)$, and using  Eqs.~(\ref{eq:e-capture-gibbs-approx}), (\ref{eq:def-F}), (\ref{eq:muebeta}), and (\ref{eq:electron-discontinuity}), we finally obtain
\begin{eqnarray}
\label{eq:heat-ocrust}
\mathcal{Q}(A,Z) &=& Q_{\rm EC}(A,Z-1)-Q_{\rm EC}(A,Z) \nonumber \\
&&- m_e c^2 \frac{C \alpha x_r}{(3\pi^2)^{1/3}}\biggl[Z^{5/3}\sigma(Z)+(Z-2)^{5/3}\sigma(Z-2)-2 (Z-1)^{5/3}\sigma(Z-1)\biggr]\, ,
\end{eqnarray}
where $x_r$ is given by Eq.~(\ref{eq:exact-xr}). The first two terms can be alternatively expressed as 
\begin{equation}
 Q_{\rm EC}(A,Z-1)-Q_{\rm EC}(A,Z) = -Q_\beta(A,Z-2) + Q_\beta(A,Z-1) = 2 \Delta^{(3)}(A,Z-1)\, , 
\end{equation}
where we have introduced the \emph{isobaric} three-point mass formula defined by 
\begin{equation}\label{eq:isobar-delta3}
 \Delta^{(3)}(A,Z) \equiv \frac{1}{2}\biggl[ 2 M^\prime(A,Z)c^2-M^\prime(A,Z+1) c^2 - M^\prime(A,Z-1)c^2 \biggr]\, .
\end{equation}
Equation~(\ref{eq:heat-ocrust}) is only valid if $g(A,Z-2,P_\beta) < g(A,Z,P_\beta)$, which is generally satisfied for 
even $A$ nuclei but not necessarily for odd $A$ nuclei. In the latter case, we typically have $Q_\beta(A,Z-1) <Q_\beta(A,Z-2)$. 
Using Eq.~(\ref{eq:muebeta}), this implies that $\mu_e^\beta(A,Z)<\mu_e^\beta(A,Z-1)$. In other words, as the pressure reaches 
$P_\beta(A,Z)$, the nucleus ($A,Z$) decays but the daughter nucleus ($A,Z-1$) is actually stable against electron capture, and therefore 
no heat is released $\mathcal{Q}(A,Z)=0$. The daughter nucleus sinks deeper in the crust and only captures a second electron in quasi equilibrium 
at pressure $P_\beta(A,Z-1)>P_\beta(A,Z)$. 

The heat $\mathcal{Q}^*(A,Z)$ released from ground-state to excited state transitions can be obtained from Eq.~(\ref{eq:heat-ocrust}) by 
substituting $M^\prime(A,Z-1)c^2$ with $M^\prime(A,Z-1)c^2+E_{\rm ex}(A,Z-1)$. Neglecting 
the small corrections due to electrostatic interactions, electron exchange and polarization effects, we find 
\begin{equation}
 \mathcal{Q}^*(A,Z) \approx \mathcal{Q}(A,Z) + 2 E_{\rm ex}(A,Z-1)>\mathcal{Q}(A,Z)\, .
\end{equation}
For odd $A$ nuclei, heat can thus only possibly be released if the first electron capture proceeds via excited states of the 
daughter nucleus, and provided 
\begin{equation}
2E_{\rm ex}(A,Z-1)-Q_\beta(A,Z-2) + Q_\beta(A,Z-1)>0\, .
\end{equation}
It should be remarked that once the excited nuclei have decayed to their ground state, and the nuclear equilibrium has been 
attained, the threshold densities and pressures between the different crustal layers will be given by Eqs.~(\ref{eq:rhobeta}) and 
(\ref{eq:Pbeta}) respectively. 

\begin{figure*}
\begin{center}
\includegraphics[scale=0.5]{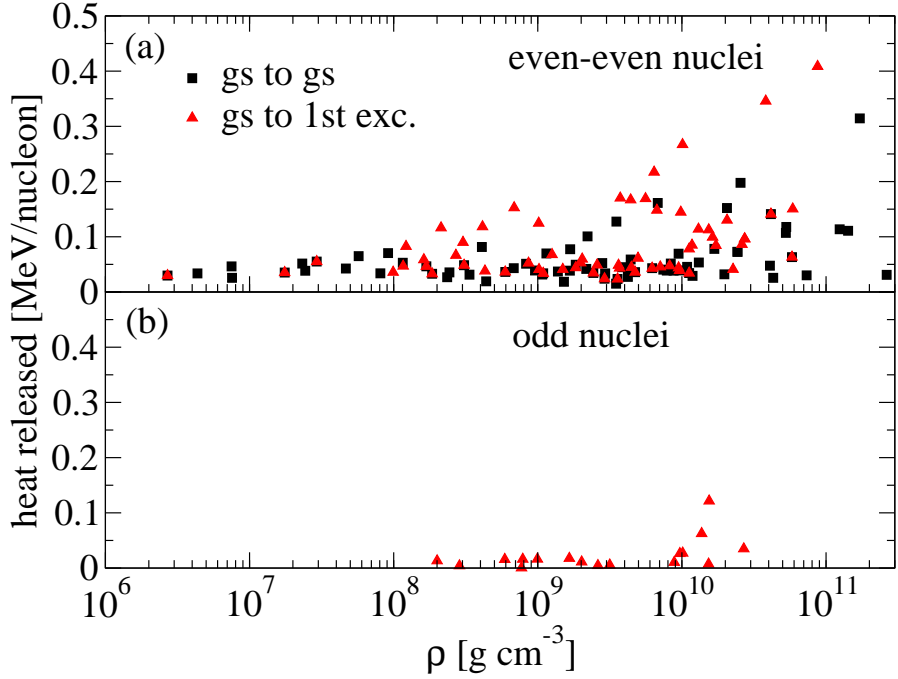}
\end{center}
\vskip -0.5cm
\caption{(Color online) 
Heat released per one accreted nucleon (in MeV) from electron captures by nuclei as a function of the average mass density (in g cm$^{-3}$) for different ashes of X-ray bursts, considering transitions from the ground state (gs) of the (a) even-even and (b) odd parent nucleus to either the ground state or the first excited (1st exc.) state of the daughter nucleus.
}
\label{fig1}
\end{figure*}

\begin{figure*}
	\begin{center}
		\includegraphics[scale=0.5]{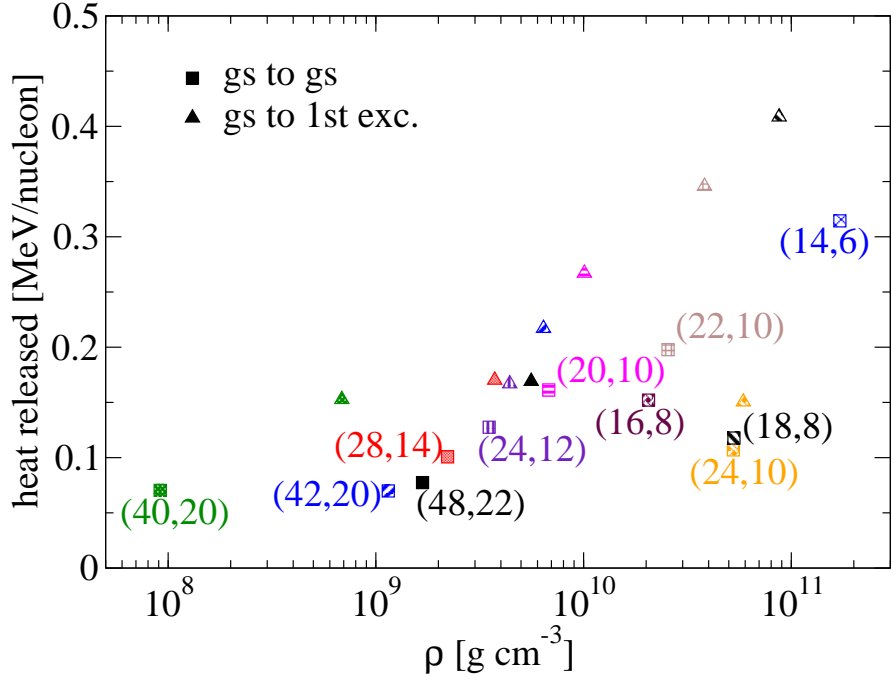}
	\end{center}
	\vskip -0.5cm
	\caption{(Color online) Same as Fig.~\ref{fig1} for the main heat sources. Numbers in parentheses refer to the mass and atomic numbers of the parent nuclei respectively. 
	}
	\label{fig2}
\end{figure*}

In the deep regions of the crust, light nuclei may undergo pycnonuclear fusion reactions at some pressure $P_{\rm pyc}$. The daughter 
nuclei are usually highly unstable against electron captures at that pressure, and thus decay by releasing some additional heat provided $P_\beta(2A,2Z)<P_\beta(A,Z)$ or $P^*_\beta(2A,2Z)<P^*_\beta(A,Z)$ depending on whether transitions 
to ground state or excited state are considered. An upper limit 
on the heat released per one accreted nucleon can be obtained by setting $P_{\rm pyc}=P_\beta(A,Z)$ or $P_{\rm pyc}=P^*_\beta(A,Z)$ respectively: 
\begin{equation}
q_{\rm pyc}(A,Z)=g(A,Z,P_\beta)-g(2A,2Z-2,P_\beta)\, ,
\end{equation}
or 
\begin{equation}
q^*_{\rm pyc}(A,Z)=g(A,Z,P^*_\beta)-g(2A,2Z-2,P^*_\beta)\, .
\end{equation}

\begin{table*}[t]
\caption{Maximum possible heat released $q_{\rm pyc}$ per one accreted nucleon from 
pycnonuclear fusion followed by electron captures 
in the crust of an accreting neutron stars. The numbers in parentheses indicate the contributions from the sole fusion. The pressure $P_{\rm pyc}$ and density $\rho_{\rm pyc}$  at which fusion occurs were fixed by the onset of electron captures considering ground-state to ground-state transitions (first two lines) and ground-state to excited state transitions (last line).
}
\label{tab:pycno}
\begin{center}
\begin{tabular}{llll}
\hline \hline \noalign{\smallskip} $P_{\rm pyc}$      & $\rho_{\rm pyc}$  & Reactions &    $q_{\rm pyc}$ 
                         \\       (dyn~cm$^{-2}$)     & (g~cm$^{-3}$)     &           &   (MeV)  \\ \hline \noalign{\smallskip}
$6.97\times 10^{28}$ & $4.15\times 10^{10}$ & $^{12}\mathrm{C} + ^{12}\mathrm{C} \rightarrow ^{24}\mathrm{Ne} -2e^-+2\nu_e $ &  1.41 (0.64) \\
$2.72\times 10^{28}$ & $2.05\times 10^{10}$ & $^{16}\mathrm{O} + ^{16}\mathrm{O} \rightarrow ^{32}\mathrm{Si} -2e^-+2\nu_e $ &  1.16 (0.58) \\
\hline 
$7.96\times 10^{28}$ & $4.59\times 10^{10}$ & $^{16}\mathrm{O} + ^{16}\mathrm{O} \rightarrow ^{32}\mathrm{Si} -2e^-+2\nu_e $ &  1.39 (0.60) \\
\hline
\hline
\end{tabular}
\end{center}
\end{table*}

\subsection{Experimental constraints}
\label{sec:results}

We have estimated the amount of heat deposited in the outer crust of accreting neutron stars and their location using the analytical formulas presented in the previous sections with the available experimental nuclear data. The \emph{nuclear} masses $M^\prime(A,Z)$ were obtained from measured \emph{atomic} masses, as compiled in the Atomic Mass Evaluation~\cite{ame16}, after subtracting out the binding energy of atomic electrons according to Eq.~(A4) in Ref.~\cite{lpt03}. Excitation energies were taken from Ref.~\cite{nds-iaea}. As for the initial composition of the ashes,
we considered two different scenarios: first, the ashes are produced by an $rp$-process during an X-ray burst~\cite{schatz2001,cyburt2016}, and second, the
ashes are produced by steady state hydrogen and helium burning~\cite{schatz2003} as expected to occur during superbursts~\cite{gupta2007}. Heat sources associated with electron captures are plotted in Fig.~\ref{fig1} as a function of the average mass density $\rho=\bar n m_u$, where $m_u$ denotes the unified atomic mass unit. We have checked that no neutron emission occurs for the transitions that are shown. Most of the heating is concentrated in the layers at densities $10^{10}-10^{11}$~g~cm$^{-3}$. As shown in Fig.~\ref{fig2}, these sources release about an order of magnitude more heat than those previously calculated in Refs.~\cite{haensel1990,haensel2003,haensel2008,fantina2018} considering ashes made of pure $^{56}$Fe or $^{106}$Pd. In particular, the capture of electrons by $^{18}$O releases about $0.4$ MeV per one accreted nucleon. 
As expected, transitions from the ground state of the parent nucleus to the first excited state of the daughter nucleus are more exothermic than ground-state to ground-state transitions. Moreover, the heat released from electron captures by odd nuclei is negligible. Pycnonuclear fusion reactions of carbon or oxygen followed by electron captures could be the main heat sources, as can be seen in Table~\ref{tab:pycno}. However, the actual amount of heat deposited depends on the abundances of these nuclei and on the unknown density at which these reactions occurs, which in turn reflect our lack of knowledge of reaction rates~\cite{yak2006,cyburt2016}. To assess the validity of the analytical treatment, we have solved numerically the threshold conditions  $g(A,Z,n_e)=g(A,Z-1,n_{e1})$ and $P(Z,n_e)=P(Z-1,n_{e1})=P(Z-2,n_{e2})$ without any approximation. The relative deviations for the transition pressure and the densities are typically of order $10^{-3}\%$, one or two orders of magnitude less for the amount of heat. For instance, considering the canonical ground-state to ground-state transition from $^{56}$Fe to $^{56}$Cr, results obtained from Eqs.~(\ref{eq:exact-Pbeta}), (\ref{eq:exact-rhobeta}), and (\ref{eq:heat-ocrust}) differ from the exact values by $-1.2\times 10^{-3}\%$, $-9.1\times10^{-4}\%$ and $3.5\times 10^{-4}\%$ respectively.

\section{Conclusions}
\label{sec:conclusion}

The heat released from electron captures and pycnonuclear reactions 
in the outer crust of accreting  neutron stars is very sensitive to the structure and excitation spectra of exotic nuclei. In particular, we have shown that the maximum amount of heat is completely determined by the isobaric three-point mass formula~(\ref{eq:isobar-delta3}), whereas their location depends on the $Q$-values of the reactions (with excitation energies suitably added to nuclear masses whenever the daughter nuclei are in an excited state). We have derived accurate analytical formulas for the pressure $P_\beta(A,Z)$ and mean baryon density $\bar{n}_\beta(A,Z)$ associated with the onset of the first electron capture by nuclei ($A,Z$), as well as the heat $q(A,Z)$ deposited.

Using the available experimental nuclear data and considering various ashes of X-ray bursts, we have found that the amount of heat deposited in the outer crust by individual sources can be up to about an order of magnitude larger than previously estimated in Refs.~\cite{haensel1990,haensel2003,haensel2008} considering ashes made of pure $^{56}$Fe or $^{106}$Pd, and whose results have been widely used in neutron-star cooling simulations. 
The main heat sources are located at densities $10^{10}-10^{11}$~g~cm$^{-3}$. Electron captures and  pycnonuclear fusions of carbon or oxygen could thus potentially explain the origin of the shallow heating introduced in cooling simulations to reproduce the observed thermal relaxation of quasipersistent SXTs. However, our calculations yield upper limits. The amount of heat effectively deposited in the outer crust may actually be lower. It depends on the uncertain initial composition and abundances of X-ray burst ashes~\cite{cyburt2016}, on the fraction of energy carried away by neutrinos produced by electron captures, and on the uncertain rates of pycnonuclear reactions~\cite{yak2006}. In any case, the large amount of heat required for interpreting the cooling data of Aql X-1 and MAXI J0556$-$332 remains a puzzle.

\begin{acknowledgments}
The work of N.C. was financially supported by Fonds de la Recherche Scientifique (Belgium) under grant No. IISN 4.4502.19. This work was also partially supported by the Centre National de la Recherche Scientifique under grant No. PICS07889 (France), the National Science Centre under grant No. 2018/29/B/ST9/02013 (Poland), and the European Cooperation in Science and Technology Action CA16214 (EU). \end{acknowledgments}


\begin{thebibliography}{99}
\bibitem{parikh2013} A. Parikh et al., Prog. Part. Nucl. Phys. 69, 225 (2013).
\bibitem{galloway} D. K. Galloway \& L. Keek, preprint, arXiv:1712.06227
\bibitem{degenaar2018} N. Degenaar \& V. F. Suleimanov in {\it The Physics and Astrophysics
of Neutron Stars}, Astrophysics and Space Science Library,  Vol. 457, edited by L. Rezzolla,  P. Pizzochero,  D. I. Jones,  N. Rea, and I. Vida\~na (Springer, Cham, 2018) p. 185. 
\bibitem{meisel2018} Z. Meisel et al., J. Phys. G: Nucl. Part. Phys. {\bf 45}, 093001 (2018). 
\bibitem{lau2018} R. Lau, M. Beard, S.S. Gupta, H. Schatz, A.V. Afanasjev, E.F. Brown, A. Deibel, L.R. Gasques, G.W. Hitt, W.R. Hix, L. Keek, P. M\"oller, P.S. Shternin, A.W. Steiner, M. Wiescher, Y. Xu, Astrophys. J.{\bf 859}, 62 (2018). 
\bibitem{fantina2018} A.F. Fantina, J.L. Zdunik, N. Chamel, J.M. Pearson, P. Haensel, S. Goriely, 
Astron. Astrophys.  {\bf 620}, A105 (2018). 
\bibitem{shchechilin2019} N. N. Shchechilin \& A.I. Chugunov, Mon. Not. R. Astron. Soc. {\bf 490}, 3454 (2019).
\bibitem{lrr} N. Chamel \& P. Haensel, Living Reviews in Relativity {\bf 11}, 10 (2008). \url{https://doi.org/10.12942/lrr-2008-10}
\bibitem{wijnands2017} R. Wijnands, N. Degenaar, D. Page, J. Astrophys. Astr. {\bf 38}, 49 (2017).
\bibitem{brown2009} E. F. Brown \& A. Cumming, Astrophys. J. {\bf 698}, 1020 (2009).
\bibitem{degenaar2011} N. Degenaar et al., Mon. Not. R. Astron. Soc. {\bf 414}, L50 (2011). 
\bibitem{degenaar2013} N. Degenaar et al., Astrophys. J. {\bf 775}, 48 (2013).
\bibitem{degenaar2014} N. Degenaar et al., Astrophys. J. {\bf 791}, 47 (2014).
\bibitem{homan2014} J. Homan et al., Astrophys. J.{\bf 795}, 131 (2014).
\bibitem{degenaar2015} N. Degenaar et al., Mon. Not. R. Astron. Soc. {\bf 451}, 2071 (2015).
\bibitem{turlione2015} A. Turlione et al., Astron. Astrophys.  {\bf 577}, A5 (2015).
\bibitem{deibel2015} A. Deibel et al., Astrophys. J. {\bf 809}, L31 (2015).
\bibitem{merritt2016} R. L. Merritt et al., Astrophys. J. {\bf 833}, 186 (2016).
\bibitem{ootes2016} L.S. Ootes et al., Mon. Not. R. Astron. Soc. {\bf 461}, 4400 (2016).
\bibitem{waterhouse2016} A. C. Waterhouse et al., Mon. Not. R. Astron. Soc. {\bf 456}, 4001 (2016).
\bibitem{degenaar2017} N. Degenaar et al., Mon. Not. R. Astron. Soc. {\bf 465}, L10 (2017). 
\bibitem{parikh2017} A.S. Parikh, et al., Astrophys. J. {\bf 851}, L28 (2017).
\bibitem{parikh2018} A.S. Parikh et al., Mon. Not. R. Astron. Soc. {\bf 476}, 2230 (2018).
\bibitem{ootes2018} L.S. Ootes et al., Mon. Not. R. Astron. Soc. {\bf 477}, 2900 (2018).
\bibitem{ootes2019} L.S. Ootes, Mon. Not. R. Astron. Soc. {\bf 487}, 1447 (2019).
\bibitem{degenaar2019} N. Degenaar et al., Mon. Not. R. Astron. Soc. {\bf 488}, 4477 (2019).
\bibitem{parikh2019} A.S. Parikh et al., Astron. Astrophys.  {\bf 624}, A84 (2019).
\bibitem{haensel2007} P. Haensel, A.~Y. Potekhin, and D.~G. Yakovlev, \textit{Neutron Stars 1: Equation of state  and structure} (Springer, Berlin, 2007).
\bibitem{medin2011} Z. Medin \& A. Cumming, Astrophys. J. {\bf 730}, 97 (2011).
\bibitem{medin2014} Z. Medin \& A. Cumming, Astrophys. J. {\bf 783}, L3 (2014).	
\bibitem{medin2015} Z. Medin \& A. Cumming, Astrophys. J. {\bf 802}, 29 (2015).
\bibitem{inogamov2010} N.A. Inogamov \& R.A. Sunyaev, Astron. Lett. {\bf 36}, 848 (2010).
\bibitem{haensel2008} P. Haensel \& J.L. Zdunik, Astron. Astrophys.  {\bf 480}, 459 (2008).
\bibitem{tondeur1971} F. Tondeur, Astron. Astrophys. {\bf 14}, 451 (1971).
\bibitem{bps1971} G. Baym, C. Pethick, and P. Sutherland, Astrophys. J. {\bf 170}, 299 (1971).
\bibitem{horowitz2008} C.J. Horowitz, H. Dussan, D.K. Berry, Phys. Rev. C{\bf 77}, 045807 (2008).
\bibitem{chamel2015} N. Chamel, A.F. Fantina, J.L. Zdunik, P. Haensel, Phys. Rev. C {\bf 91}, 055803 (2015). 
\bibitem{fantina2016a} A. F. Fantina, N. Chamel, Y. D. Mutafchieva, Zh. K. Stoyanov, L. M. Mihailov, and R. L. Pavlov, Phys. Rev. C {\bf 93}, 015801 (2016). 
\bibitem{chugunov2019} A.I. Chugunov, Mon. Not. R. Astron. Soc. {\bf 483}, L47 (2019). 
\bibitem{baiko09} D.A. Baiko, Phys. Rev. E {\bf 80}, 046405 (2009). 
\bibitem{salpeter54} E. E. Salpeter, Australian J. Phys. {\bf 7}, 373 (1954). 
\bibitem{cf16} N. Chamel, A.F. Fantina, Phys. Rev. D {\bf 93}, 063001 (2016). 
\bibitem{pot00} A.Y. Potekhin, G. Chabrier, Phys. Rev. E{\bf 62}, 8554 (2000). 
\bibitem{pearson2011} J.M. Pearson, S. Goriely, N. Chamel, Phys. Rev. C{\bf 83}, 065810 (2011). 
\bibitem{chamel2016} N. Chamel, A.F. Fantina, Phys. Rev. C {\bf 94}, 065802 (2016). 
\bibitem{chamel2020} N. Chamel, Phys. Rev. C {\bf 101}, 032801(R) (2020). 
\bibitem{BisnoSeidov1970} G.S. Bisnovatyi-Kogan \&  Z.F. Seidov, Sov. Astronomy {\bf 14}, 113 (1970).
\bibitem{BisnoBook2001} G.S. Bisnovatyi-Kogan,  \textit{Stellar Physics 1: Fundamental Concepts and 
Stellar Equilibrium} (Springer, 2001), pp.223-225.
\bibitem{ame16} M. Wang, G. Audi, F.~G. Kondev, W.~J. Huang, S. Naimi, X. Xu, Chinese Physics C {\bf 41}, 030003 (2017). 
\bibitem{lpt03} D. Lunney, J.M. Pearson, C. Thibault, Rev. Mod. Phys. {\bf 75}, 1021 (2003). 
\bibitem{nds-iaea} \url{https://www-nds.iaea.org/relnsd/NdsEnsdf/QueryForm.html}
\bibitem{schatz2001} H. Schatz et al. Phys. Rev. Lett. {\bf 86}, 3471 (2001). 
\bibitem{cyburt2016} R. H. Cyburt et al., Astrophys. J. {\bf 830}, 55 (2016). 
\bibitem{schatz2003} H. Schatz, L. Bildsten, A. Cumming, M. Ouellette, Nucl. Phys. {\bf A718}, 247c (2003). 
\bibitem{gupta2007} S. Gupta, E.F. Brown, H. Schatz, P. M\"oller, K.L. Kratz, Astrophys. J. {\bf 662}, 1188 (2007). 
\bibitem{haensel1990} P. Haensel \& J.L. Zdunik, Astron. Astrophys.  {\bf 227}, 431 (1990). 
\bibitem{haensel2003} P. Haensel \& J.L. Zdunik, Astron. Astrophys.  {\bf 404}, L33 (2003). 
\bibitem{yak2006} D.G. Yakovlev, L.R. Gasques, A.V. Afanasjev, M. Beard, M. Wiescher,  Phys. Rev. C{\bf 74}, 035803 (2006). 
\end{thebibliography}
\end{document}